# LTE Spectrum Sharing Research Testbed: Integrated Hardware, Software, Network and Data*

Vuk Marojevic
Bradley Department of Electrical and Computer Engineering
Virginia Tech, Blacksburg, VA, USA
maroje@vt.edu

Randall Nealy
Bradley Department of Electrical and Computer Engineering
Virginia Tech, Blacksburg, VA, USA
rnealy@vt.edu

Jeffrey H. Reed
Bradley Department of Electrical and Computer Engineering
Virginia Tech, Blacksburg, VA, USA
reedjh@vt.edu

## ABSTRACT

This [1] paper presents Virginia Tech's wireless testbed supporting research on long-term evolution (LTE) signaling and radio frequency (RF) spectrum coexistence. LTE is continuously refined and new features released. As the communications contexts for LTE expand, new research problems arise and include operation in harsh RF signaling environments and coexistence with other radios. Our testbed provides an integrated research tool for investigating these and other research problems; it allows analyzing the severity of the problem, designing and rapidly prototyping solutions, and assessing them with standard-compliant equipment and test procedures. The modular testbed integrates general-purpose software-defined radio hardware, LTE-specific test equipment, RF components, free open-source and commercial LTE software, a configurable RF network and recorded radar waveform samples. It supports RF channel emulated and over-the-air radiated modes. The testbed can be remotely accessed and configured. An RF switching network allows for designing many different experiments that can involve a variety of real and virtual radios with support for multiple-input multiple-output (MIMO) antenna operation. We present the testbed, the research it has enabled and some valuable lessons that we learned and that may help designing, developing, and operating future wireless testbeds.

## 1 INTRODUCTION

Long-term evolution (LTE) has become the standard for fourth generation (4G) mobile communications. LTE has been deployed in many parts of the world and allows true broadband wireless communications services on the move. Whereas the fundamental concepts of LTE, as standardized by the 3rd Generation Partnership Project (3GPP) in Release 8 (Rel. 8), are well understood, its massive deployment, emerging applications and new features bring along new research and development (R&D) challenges. For example, 3GPP is finalizing Rel. 14, which adds new cellular vehicle to everything (C-V2X) capabilities, among others. Some of the R&D challenges of C-V2X are reliability and capacity. Release 13 introduced LTE operation in unlicensed radio frequency (RF) spectrum. Licensed assisted access (LAA) uses carrier aggregation, supported since Rel. 10, to aggregate an LTE channel in the unlicensed spectrum to a primary channel in a licensed LTE band. LAA or LTE-Unlicensed (LTE-U) needs to coexist with WiFi and legacy radios.

In the US, the First Responder Network Authority (FirstNet) has been established to provide emergency responders with the first nationwide broadband public safety network using LTE Rel. 8 or higher. The deployment of such network is underway. The National Institute of Standards and Technology (NIST) recently announced 33 R&D projects to help accelerate successful deployment. The emphasis areas include LTE open-source software and testbed development using software-defined radio (SDR) technology.

Testbeds play a major role in developing and testing new wireless communications technologies and systems. They enable research and education on various aspects of radio communications system design, deployment, operation and evolution. This paper presents the evolution of our LTE testbed presented in [1]. The testbed extended RF network allows experimenting with advanced communications features, including multiple-input multiple-output (MIMO) communications and Narrowband Internet of Things (NB-IoT). The revised testbed supports operation over extended frequency bands. In addition, we offer new data that represent samples of a radio signal and allow recreating a real radio environment for spectrum sharing experiments. An open-source flexible OFDM waveform generator targeting LTE research has been implemented and can be configured to recreate one or more LTE control channels or rapidly implement and test new channels. The core of this versatile testbed is integrated in a full-size rack and features several LTE base stations or eNodeBs (eNBs), user equipment (UEs), RF channel emulators, a configurable RF network, software and data supporting research in a controlled laboratory setup. This, together with distributed SDR/RF components and hooks for adding additional equipment, provides a unique wireless communications testbed for rapid prototyping and quick evaluation of research contributions to the latest 3GPP standard specifications.

Reference [1] provides a comprehensive overview of university LTE testbeds and their capabilities. These include the LTE-A testbed [2] at TU Dresden, Germany, the UC4G wireless MIMO testbed [3] at Heriott-Watt University in Edinburg, UK, NITOS [4] at University of Thessaly, Greece, Cloud radio access network (CRAN) research testbed [5] at Campus Universitario de Santiago, Portugal, and ORBIT [6] at Rutgers University, USA. This testbed is unique in several ways, including

- Channel emulated and radiated RF signaling,
- Modular design, easily extensible and upgradable,
- Custom software and recorded radio samples,
- FCC experimental license,
- Industry-grade LTE test equipment, and
- Controlled RF environment enabling jamming and spectrum sharing experiments, among others.

The rest of the paper is organized as follows. Section 2 provides an overview, before discussing the hardware, software, network and data in detail in Sections 3-6. Section 7 discusses some enabled research. Sections 8 and 9 describe some of the key lessons learned and conclude the paper with an outlook.

## 2  SYSTEM OVERVIEW

The testbed is built of general and special-purpose processing hardware, software, and a flexible Ethernet and RF network. The networked testbed allows remotely configuring the system and experiment. It features open-source SDRs, commercial software and hardware, and industry-grade LTE test equipment.

Fig. 1 shows a block diagram of the testbed's main hardware components and RF connections. The RF signals can access the wireless channel through eight multipurpose RF ports. Five multiband antennas are currently deployed in the ceiling of our RF laboratory (lab) with three additional ports for connecting other antennas, lab or user-provided equipment to the testbed. Alternatively, the signals can be routed through configurable RF channel emulators. This allows for non-radiating experiments in a controlled RF environment. The testbed features different implementations of LTE base stations (eNodeBs or eNB) with their evolved packet cores (EPCs). These are available as SDR implementations or hardware emulators. In addition to SDR UEs, several commercial UEs of different categories and types are available. A shielded box can be used for controlled over-the-air experiments and the required RF isolation in such a testbed. An FCC experimental license for several bands is also in place. Table 1 summarizes the testbed's main features.

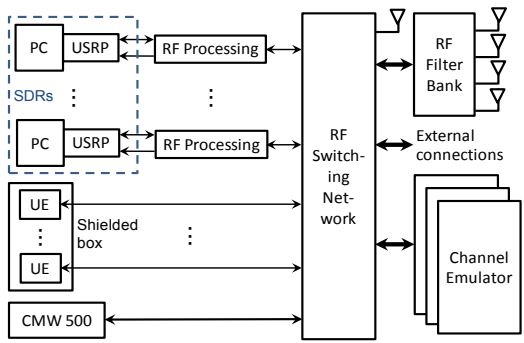

**Figure 1: Block diagram of main hardware components with simplified RF network.**

**Table 1: LTE spectrum sharing research testbed features**

| Feature | Support |
|---|---|
| Standards | 3GPP LTE Rel. 8-13 |
| | IEEE 802.11a,g,n |
| | Customizable synthetic and real waveforms |
| Frequencies | Hardware supports up to 6 GHz |
| | Antennas: 698-960, 1710-2700, 2700-3200 MHz |
| Licenses | FCC experimental license, several bands |
| | between 450 and 3650 MHz [7] |
| Channels | Channel emulation |
| | Over-the-air transmission |
| Reference signals | 10 MHz and pulse per second (PPS) references (OctoClock) |
| Network | Configurable RF network, Ethernet network |
| Software | Testbed configuration and SDR software |
| Data | Sampled radar signals |
| Access | Physical and remote |

Our design uses an RF switching network to select individual device ports for interfacing with other devices. The entire testbed is remotely accessible and configurable through the Internet. The IP-based Ethernet network allows accessing individual equipment, including SDRs, RF switches, channel emulators and other instruments.

## 3  HARDWARE

Fig. 2 shows a photo of the main equipment rack. Its components are described in continuation.

### 3.1  LTE eNB Emulator and UE Tester

The CMW500 from Rohde & Schwarz is a wideband communication system tester. Our CMW500 is currently equipped with FD-LTE and TD-LTE signaling that is compliant with 3GPP Rel. 8. It allows LTE signaling in any arbitrary RF band below 6 GHz. As an LTE UE test instrument, it allows the monitoring of LTE performance parameters, such as throughput, block error rate (BLER), and channel quality indicator (CQI), in real-time. Data logging is available for offline analysis. The CMW500 can also serve as a spectrum analyzer (SA) and provide WiFi or 3G signaling. Its main screen can be exported and all knobs and meters remotely accessed (Fig. 3). Additional 3GPP or other functionalities can be added any time through software, hardware of firmware upgrades.

### 3.2  Software-Defined Radio Hardware

An SDR hardware system consists of a computer and one or more USRPs. The computing nodes provide the software processing capabilities of the testbed. An 18-core, Dual Intel Xeon Processor E5-2694 v4 embedded in a rack-mountable workstation allows concurrently running several waveforms. Three high-performance desktops are shown on the bottom of Fig. 2. The mobile workstations (8 high-performance laptops) can be connected to the internal network of the fixed infrastructure or used individually.

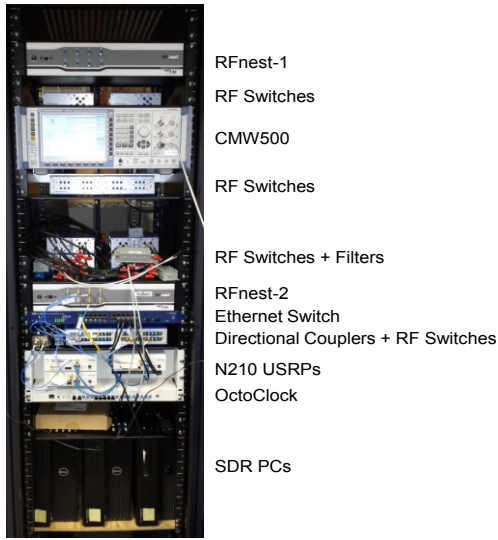

Figure 2: **Main equipment rack (cabling partially shown for clarity).**

Three Intel i5/i7 NUC mini PCs serve dual purposes: (1) mobile SDR computing nodes and (2) hubs for powering, controlling, and accessing the LTE UEs (Section 4).

The USRPs that we use in this testbed are of type N210, B210, and E310. The N210 is connected through 1000BaseT Ethernet to a computer, the B210 uses USB 3.0, whereas the E310 operates without a computer. Each N210 has a modular RF daughterboard of type SBX. This board covers frequencies between 440 and 4400 MHz and support an instantaneous bandwidth of 40 MHz. The B210s operate at 0.1–6 GHz with 60 MHz instantaneous bandwidth. The E310s have similar specifications.

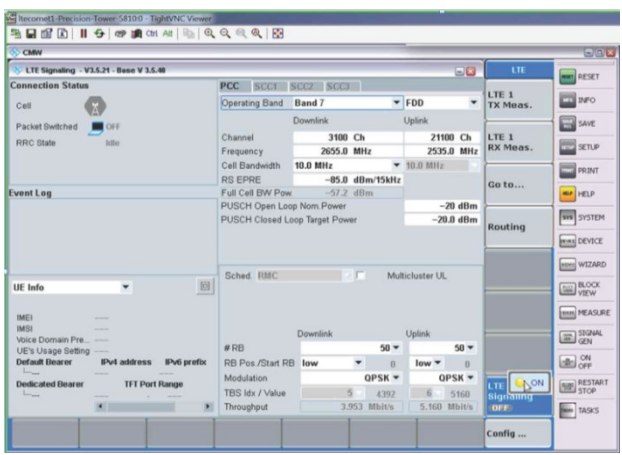

Figure 3: **CMW500 remote access control panel.**

The testbed has three N210s, which are integrated into the main rack above the PCs in Fig. 2, three E310s, which only need a power source, and ten B210s, which can be connected to desktops, laptops or the rackmount workstation. The B210s are used primarily with the portable nodes and a single laptop can connect to several B210s.

### 3.3 RF Processing Blocks

Each N210 USRP has an RF processing block attached to it (Fig. 1). This processing consists of RF attenuators and combiners. The combiner takes the form of a wideband 10 dB directional coupler that allows both USRP ports—the receive only port and the transmit/receive port—to be used. This configuration provides 10 dB of attenuation to the transmitted signal while allowing for reception with no additional attenuation. An additional 10 dB attenuator is provided at the channel emulator port to reduce the transmitted signal by a total of 20 dB. Since the USRP N210 with SBX daughterboard can output 100 mW, the system provides 0 dBm maximum signal strength to the channel emulator port and 10 dBm to the RF filter bank.

The combiners are primarily used to facilitate full duplex operation (on different frequencies). A single SDR can then implement an FD-LTE eNB. Without the combiner, two RF channel emulator ports would be required. The attenuators are needed to match the signal levels of the SDRs to the channel emulator levels.

### 3.4 Channel Emulator

Intelligent Automation, Inc.'s RFnest is a RF network channel emulation and simulation tool. The channel emulator hardware applies signal attenuation between desired signal ports to create RF channels under program control. The A208 model is a digitally-controlled analog RF channel emulator that allows up to eight simultaneous RF connections. It is accessed via Ethernet.

The two RFnest channel emulators in our testbed cover different bands: 0–1 and 1.8–2.8 GHz. A third, custom-built channel emulator is being implemented and will cover a wider range of frequencies of up to 6 GHz. The eight RF ports of each A208 model can be used to establish four simultaneous single-input single-output (SISO) radio links that can be fully isolated or not, creating different radio environments with different levels of adjacent or co-channel interference. Different combinations of MIMO and SISO radios can also be deployed. Moreover, the A208 allows creating a network with physical and virtual radios.

The main limitations of the analog channel emulator are the low number of physical ports and the fact that no delays or Doppler effects can be emulated. The analog solution serves our purpose and we chose it over the more sophisticated digital RFnest [9].

## 3.5 Filters and Antennas

Since the USRP RF components of the SBX daughterboard (used with the N210), for example, only include a single fixed filter (cutoff at the highest specified daughterboard frequency of 4.4 GHz in this case), additional filtering must be provided in order to suppress transmitted harmonics and spurious receiver responses. In other words, filtering is required for (a) regulatory compliance and (b) radio performance. Configurable filter banks are therefore provided for use in conjunction with the antenna system for controlled over-the-air operation. We have four filter banks, one for each N210 and one for a B210. The pass bands are 800-1000, 2025-2075, 2350-2550 and 3550-3650 MHz, each corresponding to one of four electronic switch positions. The setup covers the US 3.5 GHz Innovation Band, which is a shared spectrum band for next generation wireless communications below 6 GHz [8]. Filters can be exchanged as needed.

The system uses five ceiling mounted radome-enclosed, omni-directional, and vertically polarized antennas operating over the frequency ranges 698-790, 790-960, 1710-2700 and 2700-3200 MHz. These antennas are located in the RF lab, which is adjacent to the server room where the main testbed rack is located.

## 3.6 User Equipment

LTE Category 4, 5 and 6 user devices in the form of USB dongles or access points are provided with the testbed: Huawei B593s-22, Huawei E3276 LTE Dongle, Huawei E8278, and Rogers Aircard U330. All but the first are USB dongles. USB dongles are powered from USB and can be accessed through USB using a browser or vendor-specific application. This allows observing the UE mode, performance and other LTE-specific radio or network parameters. An Intel NUC or a laptop is used for the purpose of interfacing with a USB dongles. The NUC can be conveniently placed in the shielded box.

A Mini-UICC Test Card from Rohde & Schwarz provides the universal subscriber identity module (USIM) for each user device. This particular card ensures smooth interfacing with the CMW500 and Amarisoft eNBs.

## 3.7 Spectrum Analyzer

The testbed includes two portable spectrum analyzers (SAs) for indoor and outdoor measurement studies over a frequency range between 10 kHz and 6.2 GHz. It is a mobile unit that can be hooked up with the testbed as needed via RF cables or antennas. Each SA has a built-in GPS receiver and supports remote operation using its Ethernet interface.

## 3.8 Reference Oscillator and Clock Source

Ettus Research Octoclock has eight 10 MHz and eight pulse per second (1 pps) reference signals. It distributes a common timing and 10 MHz reference signal to the USRPs and CMW500. The use of it is optional. You can select through the USRP hardware driver (uhd) whether to use an internal or external reference signals. Octoclock allows to provide an increase in frequency accuracy. The 1 pps signal allows the sample clocks to be aligned.

## 4 SOFTWARE

### 4.1 Wireless Communications Software

This is the main software and features LTE software as well as several toolboxes for building SDRs for diverse experiments.

*4.1.1 Amarisoft LTE100 eNB.* Amarisoft's Software eNB is installed on several workstations, on two fixed and one mobile node. It supports operation with USRPs B210 and N210. Amarisoft is able to connect to several UEs simultaneously. It is well maintained continuously upgraded with the newest 3GPP releases. The latest release features NB-IoT.

Fig. 4 shows a screenshot of the MAC trace for a 2x2 MIMO DL experiment. The first 'brate' column matches the theoretical maximum DL user throughout of 102 Mbps for a 20 MHz FD-LTE system and Cat. 3 UE. The other parameters are typical LTE link parameters. They are helpful to understand system conditions and educate students about the fundamental parameters used by eNBs to manage the uplink and downlink.

*4.1.2 Amarisoft LTE100 UE.* This software allows several UEs to be emulated and controlled through a GUI interface. More precisely, it can simulate the behavior of up to 64 LTE UEs connected to an eNB via 3GPP compliant LTE signaling. The combined signals go through a channel emulator or over-the-air. LTE100 UE is installed on the rackmount workstation because of the required processing power.

*4.1.3 srsLTE.* srsLTE is a free, open-source SDR library for implementing 3GPP compliant LTE system on general-purpose processors (GPPs). It has a modular structure with minimal inter-modular and external dependencies. The current version is compliant with LTE Release 8, and is written entirely using C language. It supports USRPs [10].

*4.1.4 Flexible OFDM Waveform - Physical Channels.* We used srsLTE to develop a flexible waveform generator for custom RF signaling. The waveform is OFDM-based with configurable bandwidth, subcarrier spacing, subcarrier allocation (regular or irregular), duty cycle, and so forth. This allows creating non-contiguous and time-discontinuous OFDM waveforms at the granularity of a single resource element (the smallest OFDM resource that carries one modulation symbol). Using the LTE resource mapper, the generated RF frame can mimic LTE physical channels, such as the LTE synchronization signals or the broadcast channel. Arbitrary physical channels can also be defined (Fig. 5).

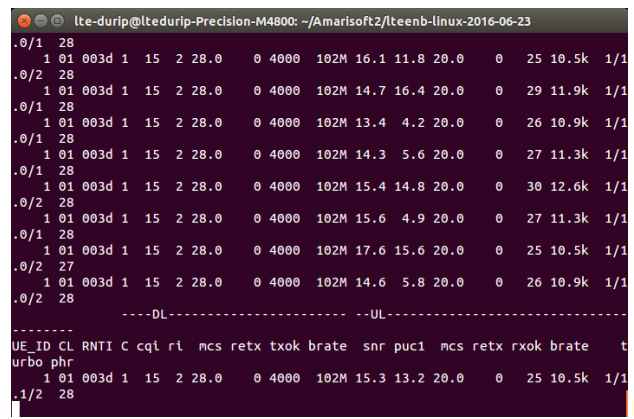

**Figure 4: Amarisoft LTE100 eNB's MAC trace for over-the-air DL measurement with 20 MHz BW and 2x2 MIMO.**

*4.1.5 GNU Radio.* GNU Radio is a free and open-source software development kit meant for implementing and rapid prototyping of DSP algorithms for SDR. It can be used with a) low-cost external RF hardware to create a software radio, or b) used without any external hardware in a purely simulation-based setting.

GNU Radio companion (GRC) is the graphical user interface for GNU Radio. It is installed on most workstations. To install GNU Radio, it is recommended to use a UHD version that is compatible with both Amarisoft and GNU Radio. UHD version 3.8.1 has been found to be compatible with all other software tools.

GNU Radio version 3.6, including GNU Radio Companion (GRC), is currently installed on most nodes. Upgrades to newer versions are possible. The mobile workstations run the latest GNU Radio version, version 3.7.

## 4.2 Channel Emulation Software

RFnest comes with software that allows controlling the channel attenuation on the different RF paths:

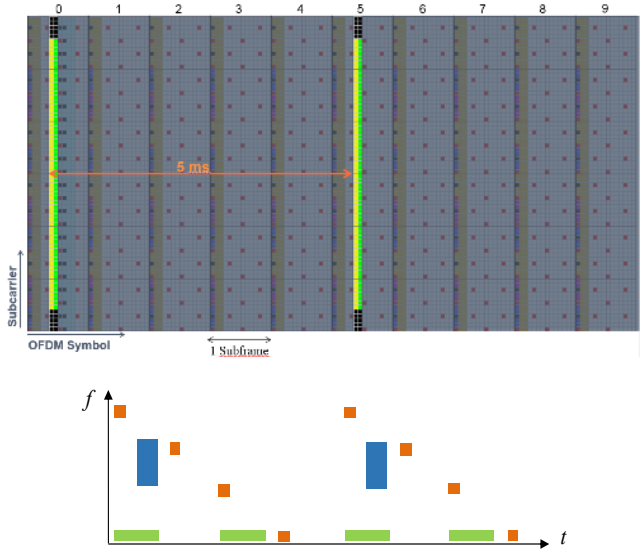

**Figure 5: OFDM resource grids: The flexible waveform can generate the LTE synchronization signals (top) or any LTE control channel or arbitrary signal (bottom).**

- **RFview GUI:** The graphical user-interface (GUI) allows for scenario modeling, analysis, recording and replay. The GUI provides time-synchronized and geospatial displays of the scenario state.
- **Channel Emulation Controller (CEC):** The CEC coordinates with RFview and carries out initialization and updates the channel emulation properties over time according to the RF scenario.

The system supports various standard propagation models and can emulate fading channels. One can specify terrain, mobility paths in three dimensions and velocities for several radios.

## 4.3 Performance Analysis Software

Performance analysis software is useful for evaluating the performance of LTE or other radios in different RF environments.

*4.3.1 iPerf and jPerf.* iPerf is a measurement tool that creates streams of traffic data and determines the maximum achievable bandwidth on IP networks.

jPerf is a GUI front-end developed in Java for iPerf. It provides an interface to select various options, which are ultimately translated to a command line interface. Both, iPerf and jPerf, are available for download on the Internet.

*4.3.2 UE Monitoring.* All commercial user devices come with an application to monitor status of the UE and other radio link parameters, such as communications mode and data rate. Some devices also allow logging into them and looking at several radio parameters through AT commands [11].

## 5 Network

### 5.1 Ethernet Network

The testbed can be accessed remotely by authorized users over a Virtual Private Network (VPN) by using a valid .OVPN certificate that was issued by the testbed administrator. The certificate is verified by the gateway computer, and once authenticated, users are granted access to the networked system components. The user can *ssh* into each one of the computers or instruments and access the channel emulator, switches, filters, UEs, for experiment configuration and execution and runtime analysis (Fig. 6).

### 5.2 RF Network

The RF switches allow reconfiguring the testbed to operate in channel emulated or radiated mode. It also allows combining different LTE test equipment, SDRs and UEs into an experiment, limited only by the number of channel emulation or antenna ports.

For 2x2 MIMO operation, the transmitter and receiver both require two ports each to be selected simultaneously, which is equal to half the number of available ports on each RF channel emulator. Apart from the USRPs and UEs, separate connections need to be provided to connect to the CMW500, to external antennas, and to additional equipment in the RF lab.

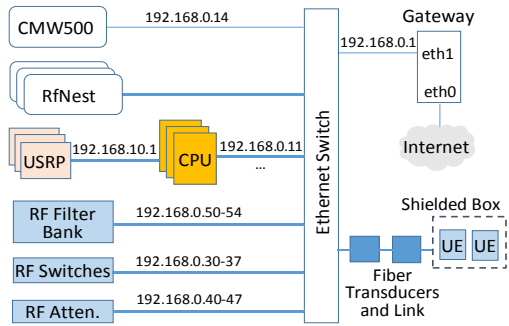

**Figure 6:** Ethernet network.

One way of accomplishing flexible configurations is through the use of a series of cascaded RF switches or switching matrix. To keep the cost low and to maximize the reuse of the original testbed setup [1], a new approach to the problem was considered where the device ports were ranked in terms of priority and several operational scenarios were considered. Combinations of these device ports were then accordingly grouped to eliminate combinations of device ports that would not be used. The selected combinations enable switching between several modes of operation with the inclusion of only one additional 8-port switch matrix. That is, three switch matrices are needed to choose among (1) over-the-air transmission, (2) RFnest1, (3) RFnest2, and (4) custom channel emulator. The forth switch matrix facilitates defining an experiment with a different combination of radios and essentially enables 2x2 MIMO. Operating at 4x4 MIMO would require additional ports on the network emulators, 4-port UEs and additional RF switching.

Fig. 7 shows the RF diagram. Switch Matrix 1 allows switching between channel emulated and radiated modes. For example, the topmost RF line from the SDRs block connects to Switch Matrix 3 and from there to one of the RF channel emulators or to the leftmost antenna of the group of four antennas. If channel emulation is chosen, Switch Matrices 3 and 4 allow choosing between one of the three channel emulators that cover different frequency ranges. All RF signals go through a channel emulator or the air interface between two antennas.

All connections from the shielded enclosure go into Switch Matrix 2, which also provides access to two of the laboratory connections. The three laboratory connections allow integrating other lab equipment or adding additional devices to the testbed. These could be UEs, SDRs, or another instruments that have RF ports (*bring your own device*). The five antennas and three lab connections are in the adjacent RF lab that students and faculty can access.

The CMW500 has direct access to a channel emulator port, an antenna port, or a lab connection.

All switches have an IP address and are electronically configurable through a browser.

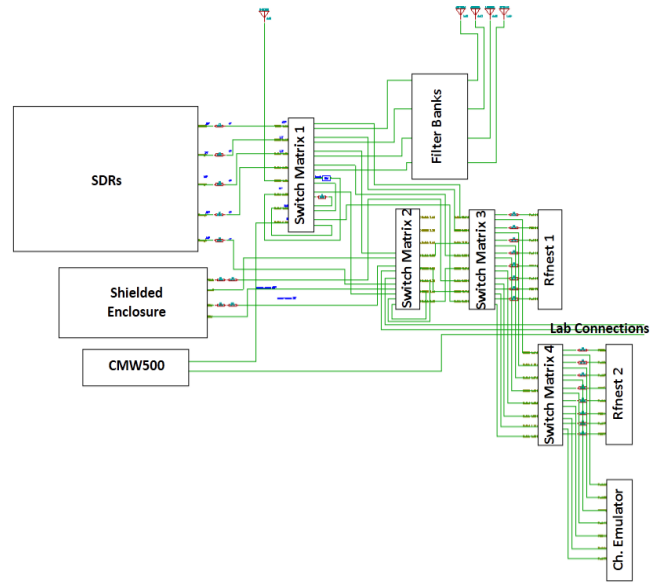

**Figure 7:** Functional RF diagram for all testbed operational modes including 2x2 MIMO.

## 6 Data

Real signals can be recorded in the field using a binary I/Q format file. The file can then be played back on one of the system USRP transmitters as either a desired signal or interferer for spectrum sharing experiments.

Several 10 minute data files were recently recorded from a live NOAA weather radar on 2.8 GHz [12]. Since the radar was scanning continuously signal levels vary widely with only occasional full strength peaks. The recordings were made using one of the portable SDR equipment with a GNU Radio flowgraph, operating at a bandwidth of 10 MHz. The resulting files are approximately 48 GB each. Since the signals were recorded as I/Q baseband data they may be replayed through a USRP transmitter at any desired center frequency. Fig. 8 shows the radar signal spectrum.

During frequency sharing experiments sample radar signal can be played back through a USRP and then directed through a channel emulator which may apply additional path loss and fading to the radar signal as appropriate for a mobile scenario. The radar signal can then be added to the communications link path of the emulator to evaluate the effects of radar interference.

## 7 ENABLED RESEARCH

The testbed enables experimental research in a controlled RF environment. Its main purpose is not only to advance LTE, but rather test and advance solutions for spectrum coexistence in heterogeneous radio environments, including LTE, WiFi and radars. We summarize some of the enabled research here and cite sources for further details and results.

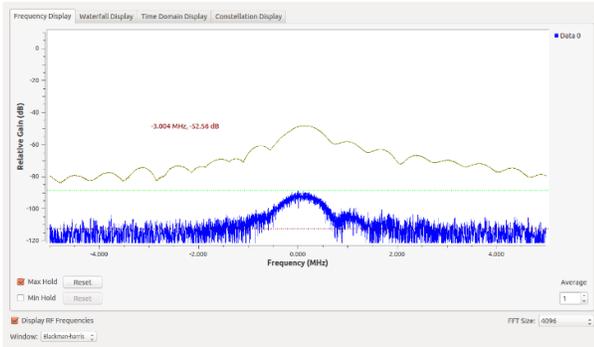

**Figure 8:** Example spectrum of the replayed radar signal. It shows a snapshot of the very wideband radar signal along with the peak hold line. The noise floor is at about -120dB.

## 7.1 Waveform Resilience to RF Interference

RF interference can be intentional or unintentional and is generated from other radios of the same type (e.g. as the result of aggressive frequency reuse in a cellular network) or different types (e.g. from other radios operating in the same band or from jammers). Our flexible OFDM waveform and SDR framework can be conveniently used to analyze the vulnerability of LTE to protocol-aware control channel interference [13]. Using Amarisoft LTE100 eNB, CMW500 and srsLTE, we have analyzed the effect of LTE control channel spoofing. The results have shown that a fake LTE downlink control signal can deceive UEs and impede their attachment to a legitimate eNB [11,14]. Solutions to this problem can be effectively implemented and tested [15].

We have also used the testbed to evaluate external LTE equipment, such as commercial, public safety, or military-grade LTE systems. The extra RF ports allows connecting external eNBs and UEs to test system performance in targeted RF interference, detect interference, and evaluate practical solutions for interference mitigation [16].

This unique system setup providing a modular and extensible framework for controlled RF experiments with commercial and experimental LTE waveforms has been a key enabler for our research on improving the availability of 4G LTE in shared spectrum and in harsh radio conditions.

## 7.2 Harmonious Spectrum Coexistence

The capacity limitations of current LTE deployments require LTE network operators to consider shared or unlicensed spectrum. Motivated by that and the success of WiFi, LTE is now considered for operation in the 5 GHz unlicensed band, the new US 3.5 GHz band, the AWS-3 band, and other shared bands around the world. The primary users in the aforementioned bands are radars and other legacy (government) users. WiFi is also present in 5 GHz. Hence, LTE needs to coexist with legacy users, WiFi, and, possibly, other radios as well.

The testbed provides WiFi signaling through the CMW500 and the UEs. Real radar signals can be emulated using our collected data, as described in Section 6. This can be effectively used to create a number of realistic scenarios that allow evaluating RF spectrum coexistence techniques.

Our testbed provides several features that allow experimental evaluation of coexistence in shared spectrum. The number of fixed and portable radio nodes allows emulating primary, secondary and tertiary users, creating a dedicated sensing network, and implementing different spectrum access rules. The *LTE user bands signaling* license of our CMW500 and capability of the deployed SDR software and hardware facilitate LTE experiments at any frequency below 6 GHz. An FCC experimental license for several bands is provided through CORNET for over-the-air experiments. The public safety LTE system described in [8] can be recreated and evaluated with additional radios, policies and coexistence mechanisms. This allows setting up different scenarios that need experimental validation of spectrum sharing concepts and rules for enabling harmonious spectrum coexistence before doing actual field trials.

## 7.3 Analysis of RF Imperfections

One emerging line of research in the context of spectrum sharing is characterizing receivers and quantifying achievable performance in heterogeneous signaling environment. As opposed to licensed spectrum, receivers will be heterogeneous, especially in multi-tier spectrum sharing scenarios. The analog components in the RF and intermediate frequency (IF) processing chains of receivers and the data converters are nonlinear devices that introduce in-band interference (intermodulation distortion or IMD) from adjacent signals. This interference can limit the achievable throughput [17], the extent of which needs to be quantified through experimental research. Receivers need to be properly characterized and this information used for dynamic spectrum allocation. This testbed is ideal for IMD testing because of the variety of signals that it can be generate within the bandwidth of the pre-selector filter. Real and synthetic signals and devices can be used to emulate various RF environments.

The goal of this ongoing research is characterizing RF imperfections and their impact on system performance. Then, these imperfections can be taken into account to adapt protocols and make better spectrum management decisions in heterogeneous and dynamic radio environments. The channel emulator, with additional RF components, can provide the necessary gain settings to drive receivers into controlled nonlinear region for validation of theoretical research results.

## 8 LESSONS LEARNED

Some of the lessoned learned are:
- **Proper RF shielding:** RF shielding is critical in such a testbeds to avoid undesired signal paths, e.g. through RF board leakage, when a cabled connection is desired. RF shielding is also important to obey the spectrum regulations and avoid commercial eNBs interfering in the experiment.
- **Remote access:** Remote access is important for avoiding physical testbed disassembly (e.g. removal of

cables) or damage. This requires extra effort for setting up the needed processes for automation, such as reimaging, to ensure the testbed is in a known state after an experiment.

- **SDR:** The use of SDR hardware and software enables versatile experiments. Open-source software libraries allow customizing the tools to enable experimental R&D with reasonable effort.
- **Automation:** Automating as many processes as possible can be of huge benefit for effective testbed operation and management. We recommend providing GUIs and scripts for improving the usability. Nevertheless, any testbed has learning curve associate with its use. We recommend that the effort be somewhat split between the manager and the user community.
- **Maintenance and upgrades:** Even if all processes are automated, there is a need for a testbed manager to perform mechanical and functional checks and replacement of parts as needed. Students who have used the testbed for research and class projects have effectively contributed to testbed maintenance and upgrades as well as training of new users. The main source of support for upgrades needs to come from research projects that leverage the existing research infrastructure or from equipment grants.
- **Compliance with specifications:** Standard compliant (commercial) components do not necessarily implement all standard features. This can lead to unexpected results. Sanity checks and revising specifications helps as does the use of alternative components. Research using a testbed like ours can convincingly point out the importance of certain features in a standard [15].

Many more lessons have been learned that have enabled research as well as education [18].

## 9 CONCLUSIONS AND OUTLOOK

This paper has introduced a testbed enabling LTE spectrum sharing and related experiments. Our testbed is unique in its components and how they are integrated. It features SDRs, industry-grade LTE test instruments and a configurable RF network, which allows connecting different components while defining an experiment. We also offer software and data that is useful for RF experiments in a controlled lab environment.

The testbed is remotely accessible and we encourage researchers and educators to contact us for use. We are planning to integrate our LTE testbed into the CORNET testbed management framework to provide integrated resource and user management. CORNET, moreover, provides tools to visualize resource status, spectrum, system performance and results in real time though a Web browser [19]. We also plan for software upgrades to support emerging 3GPP releases, that is, 4.5 and 5G technology, such as LTE-U, CRAN, and C-V2X. In the CRAN configuration, the testbed would allow executing LTE waveforms in a central location and serving remote radio heads that could be located indoors or outdoors. We also envision experiments with unmanned aerial vehicles to assess the performance of LTE-like waveforms and shared spectrum solutions for next generation unmanned aerial systems [20].

## ACKNOWLEDGMENTS

This work was supported in part by the Army Research Office contract numbers W911NF-14-1-0553/0554 and the National Science Foundation (NSF) contract number CNS-1642873. The authors would like to thank Deven Chheda, Raghunandan M. Rao and Pradeep Reddy Vaka for their contributions to the design, development and validation of the testbed. The open access of this article is facilitated by Virginia Tech's Open Access Subvention Fund.